\documentclass[
  reprint,
  superscriptaddress,
  floatfix,
  amsmath,
  amssymb,
  aps
]{revtex4-2}
\usepackage{graphicx}
\usepackage{dcolumn}
\usepackage{bm}
\begin{document}
\title{Field-Gated Programming of Nested Skyrmion Bags by Spatially Selective Skyrmion Writing}
\author{Shuhuang Lin}
\affiliation{School of Physics and Materials Science, Guangzhou University, Guangzhou 510006, China}
\affiliation{School of Physical Sciences, Great Bay University, Dongguan 523000, China}
\author{Shirong Lin}
\email{shironglin@gbu.edu.cn}
\affiliation{School of Physical Sciences, Great Bay University, Dongguan 523000, China}
\affiliation{Great Bay Institute for Advanced Study, Dongguan 523000, China}

\begin{abstract}
Nested skyrmion bags are composite magnetic solitons whose internal
occupation numbers provide an expanded state space for multilevel
information encoding. However, deterministic and region-selective
writing of individual skyrmions into nested bags remains challenging.
Here, using micromagnetic simulations, we propose a field-gated
nanocontact protocol for generating and programming double-layer
nested skyrmion bags \(S(m,S(n))\). The protocol exploits the opposite
field responses of bag regions with antiparallel magnetic
backgrounds. Spatially segmented perpendicular magnetic fields
selectively expand either the outer region or the inner bag
until it overlaps with a nanocontact. A spin-polarized current pulse
of the appropriate polarity then nucleates one skyrmion in the
selected region, while subsequent field-driven contraction
moves the written skyrmion away from the contact and restores the
writing site for repeated operation. Starting from an empty nested
bag \(S(0,S(0))\), this procedure enables the controlled construction
of \(S(m,S(0))\), \(S(0,S(n))\), and general \(S(m,S(n))\) states. The simulated total topological charge changes
in unit steps during sequential writing, confirming one-at-a-time
and region-selective control. These results provide a possible
route toward programmable nested topological states for multilevel
spintronic information encoding.

\end{abstract}

\maketitle

\section{Introduction}

Topological magnetic textures have attracted considerable interest
for robust and energy-efficient spintronic applications
\cite{nagaosa2013topological,fert2017magnetic,back20202020,
gobel2021beyond,zhang2026reversible,littlehales2026emergent,qin2026thermal,kajihara2026spin,zou2026effective,zhang2026annealing,han2026deformation,miriyala2019spin,fert2013skyrmions}. Among them, magnetic skyrmions are promising
information carriers because of their nanoscale size
\cite{wang2018theory,wu2021size}, topological
stability
\cite{leonov2016properties,wild2017entropy,oike2016interplay,bessarab2018lifetime}, and efficient current-induced manipulation
\cite{fert2013skyrmions,iwasaki2013current,juge2022skyrmions,fert2013skyrmions}. Beyond isolated
skyrmions, composite textures such as skyrmioniums
\cite{zhang2016control,powalla2023skyrmion,nakamura2024mechanism,qiu2024skyrmionium,zhang2026skyrmion}, target skyrmions
\cite{zheng2017direct,zhang2023room,cortes2019nanoscale,song2019field},
and skyrmion bags
\cite{foster2019two,kind2020existence,yang2026design,hu2026library,ghosh2026linked,zhou2026disorder,zhang2026magnetic,jiang2025stable,kern2025controlled,liu2025room,yang2024embedded,kind2021magnetic,sun2025localized,li2025observation,wang2023particle,shen2021electrically,powalla2023seeding,chen2022voltage,zeng2020dynamics,shi2025deformation,kuchkin2025symmetry} offer additional internal degrees of freedom for
information encoding. A conventional
skyrmion bag consists of a closed domain-wall-like boundary enclosing
multiple skyrmions in an oppositely magnetized local background
\cite{foster2019two,rybakov2019chiral}. The number of enclosed
skyrmions defines an occupation-number degree of freedom, providing
a natural basis for multilevel encoding
\cite{tang2021magnetic,kind2021magnetic}.

The bag concept can be extended hierarchically by embedding an
additional subbag inside the outer bag. A double-layer nested
skyrmion bag is denoted by \(S(m,S(n))\), where \(m\) is the number
of skyrmions in the outer region and \(n\) is the number
enclosed by the inner bag \(S(n)\)
\cite{foster2019two,li2024manipulating}. In contrast to a conventional
bag described by a single occupation number, a nested bag possesses
two internal occupation numbers, \(m\) and \(n\). The pair \((m,n)\)
therefore defines an expanded configurational state space. Moreover,
because the outer and inner regions possess opposite magnetic
backgrounds, the skyrmions hosted in them contribute opposite
topological charges, allowing nested bags to access positive,
negative, and compensated total topological-charge states.

Recent experiments have demonstrated magnetic skyrmion bags in
chiral magnets, van der Waals ferromagnets, and magnetic multilayers,
including embedded, room-temperature, zero-field-stable, and
site-selectively formed states
\cite{yang2024embedded,liu2025room,kern2025controlled}. Importantly,
double-nested N\'eel-type skyrmion bags have recently been observed
at room temperature in the van der Waals ferromagnet
Fe\(_{3-x}\)GaTe\(_2\), where pulsed currents and magnetic fields were
used to generate and transform different composite states
\cite{jiang2025stable}. These results establish nested skyrmion bags
as experimentally accessible magnetic textures and motivate the
development of methods for programming their internal states.

Despite these advances, independently programming the skyrmion
populations in the different regions of a nested skyrmion bag
remains challenging. For conventional non-nested skyrmion bags, an
expansion--injection--contraction protocol has been proposed, in which
a uniform perpendicular magnetic field enlarges the outer bag and a
local nanocontact nucleates the enclosed skyrmions
\cite{bo2023controllable}. This method, however, addresses only a
single bag region. For nested skyrmion bags, microwave magnetic
fields have been proposed to selectively excite and annihilate
skyrmions at different radial positions, thereby modifying the total
topological charge \cite{li2024manipulating}. Such a resonance-based
approach primarily removes pre-existing skyrmions from a prepared
nested texture. However, deterministic and region-selective addition of individual skyrmions, which is essential for independently programming \(m\) and \(n\), remains challenging. In particular, a protocol that integrates the creation of an empty nested-bag framework with subsequent, cyclically repeatable loading of skyrmions into either region is still lacking.

Here, we propose a field-gated nanocontact protocol for the
controlled generation and region-selective programming of
double-layer nested skyrmion bags \(S(m,S(n))\). The protocol first
creates an empty nested bag \(S(0,S(0))\), which provides the outer
bag and inner bag boundaries for subsequent writing. A spatially
segmented perpendicular magnetic field is then used as a magnetic
gate to selectively expand either the outer region or the inner
bag until the targeted region overlaps with the nanocontact. A
spin-polarized current pulse is subsequently applied through the
contact to nucleate one skyrmion in the selected region.
Field-driven contraction then moves the newly written skyrmion away
from the contact and restores the writing site, allowing the
expansion--writing--contraction cycle to be repeated. In this way,
\(S(m,S(0))\), \(S(0,S(n))\), and general \(S(m,S(n))\) states can
be generated in a controlled and region-selective manner.

The proposed protocol further enables programmable control of the
total topological charge. With the sign convention adopted in this
work, each skyrmion written into the outer region contributes
\(Q=-1\), whereas each skyrmion written into the inner bag
contributes \(Q=+1\). Micromagnetic simulations show that the total topological charge changes in unit steps during sequential writing,
consistent with the addition of one skyrmion to the selected
region in each successful cycle. The proposed approach
complements existing creation and annihilation protocols by
integrating the generation of an empty nested-bag framework with
one-at-a-time loading of either region, providing a possible
route toward programmable nested topological states for multilevel
spintronic encoding.

\begin{figure}
\centering
\includegraphics[width=\columnwidth]{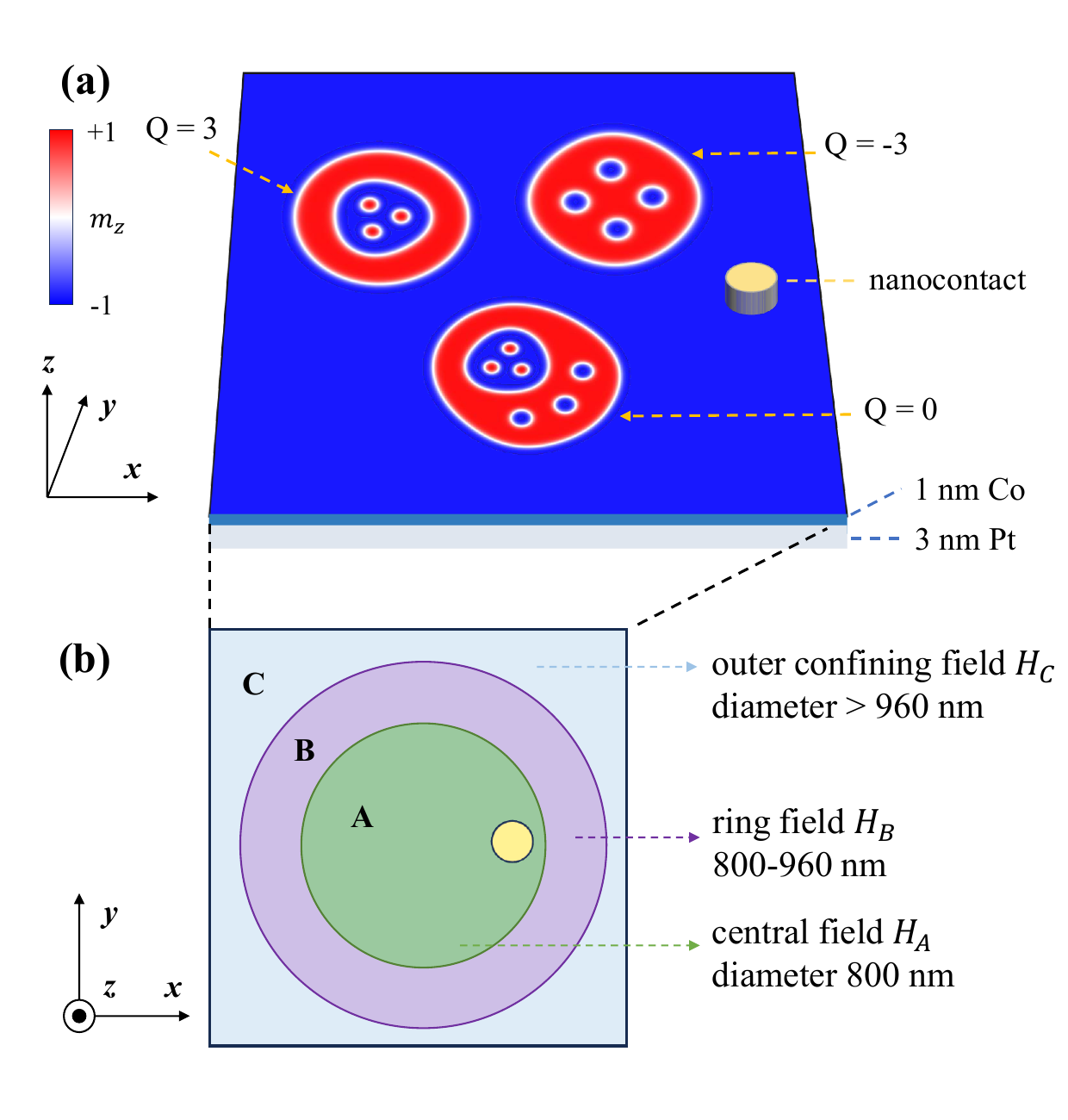}
\caption{
Schematic illustration of nested skyrmion-bag structures, the circular
nanocontact, and the spatially segmented magnetic-field regions.
(a) Representative magnetization configurations of nested skyrmion
bags with different total topological charges \(Q\) in a
Co(\(1~\mathrm{nm}\))/Pt(\(3~\mathrm{nm}\)) bilayer. The color map
denotes the normalized out-of-plane magnetization component \(m_z\),
where red and blue correspond to \(m_z=+1\) and \(m_z=-1\),
respectively. The yellow cylinder represents a circular nanocontact
with a diameter of \(d_{\rm c}=30~\mathrm{nm}\), through which a
perpendicular spin-polarized current is injected into the magnetic
layer. The resulting Slonczewski spin-transfer torque is used for
local skyrmion writing. 
(b) Spatial partition of the applied perpendicular magnetic field in
the \(1024~\mathrm{nm}\times1024~\mathrm{nm}\) square film. The film is
divided into three concentric magnetic-field regions. The
perpendicular magnetic fields in regions A, B, and C are denoted by
\(H_A\), \(H_B\), and \(H_C\), respectively.
}
\label{fig1}
\end{figure}

\section{Model and Methodology}

The topological charge of a two-dimensional spin texture is defined as \cite{nagaosa2013topological}
\begin{equation}
Q=\frac{1}{4\pi}\int \mathbf{m}\cdot
\left(
\frac{\partial \mathbf{m}}{\partial x}
\times
\frac{\partial \mathbf{m}}{\partial y}
\right) dxdy,
\label{eq:topological_charge}
\end{equation}
where \(\mathbf{m}=\mathbf{M}/M_s\) is the normalized magnetization
vector. In this work, we focus on double-layer nested skyrmion bags
denoted by \(S(m,S(n))\), as schematically shown in Fig.~\ref{fig1}(a).
Here, \(m\) represents the number of skyrmions confined in the outer bag,
whereas \(n\) denotes the number of skyrmions confined in the inner
bag \(S(n)\). Because the outer and inner bag regions possess opposite
background magnetizations, skyrmions written in these two regions
contribute opposite topological charges. With the present choice of
background magnetization and coordinate convention, each skyrmion in
the outer bag carries \(Q=-1\), whereas each skyrmion in the inner bag
carries \(Q=+1\). In the present convention, the outer and inner bag
boundaries form a compensated pair in the initial \(S(0,S(0))\) state,
yielding an approximately zero net topological charge. Therefore, the
variation of the total topological charge during the writing process is
dominated by the skyrmions nucleated inside the outer and inner bag
regions, and the total topological charge of the nested skyrmion bag is
given by
\begin{equation}
Q_{\rm total}=n-m.
\label{eq:qtotal}
\end{equation}
For example, the corresponding total topological charges of \(S(3,S(0))\),
\(S(3,S(3))\), and \(S(0,S(3))\) are \(-3\), \(0\), and \(+3\),
respectively.

We consider a chiral ferromagnetic thin film with perpendicular
magnetic anisotropy and interfacial Dzyaloshinskii--Moriya interaction
(DMI). As shown in Fig.~\ref{fig1}, the simulated system is based on a
Co(\(1~\mathrm{nm}\))/Pt(\(3~\mathrm{nm}\)) bilayer film with a lateral
size of \(1024~\mathrm{nm}\times1024~\mathrm{nm}\) and a ferromagnetic
thickness of \(1~\mathrm{nm}\). Only the Co free layer is explicitly included in the micromagnetic simulations. Periodic boundary
conditions are applied along the \(x\) direction, making the system
effectively infinite in this direction. The nanocontact, with a diameter of
\(d_{\rm c}=30~\mathrm{nm}\), is centered at
\((x_{\rm c},y_{\rm c})=(300~\mathrm{nm},0)\), where the coordinate
origin is taken at the film center.  Such a perpendicular spin-polarized current may be
provided by a vertically magnetized polarizing layer incorporated
above the nanocontact \cite{sampaio2013nucleation,yuan2016skyrmion}. The Pt layer is primarily introduced to provide
the interfacial DMI required for stabilizing N\'eel-type chiral
textures. As shown in Fig.~\ref{fig1}(b), the film is divided into three
concentric magnetic-field regions to enable region-selective
manipulation of the nested skyrmion bag.

Region A is a central circular region with a diameter of
\(800~\mathrm{nm}\). Region B is an annular region surrounding region
A, with inner and outer diameters of \(800\) and \(960~\mathrm{nm}\),
respectively. Region C is the remaining region outside the circle with
a diameter of \(960~\mathrm{nm}\). The perpendicular magnetic fields
applied in regions A, B, and C are denoted by \(H_A\), \(H_B\), and
\(H_C\), respectively. Their spatial distribution is written as
\begin{equation}
H_z(r)=
\begin{cases}
H_A, & 0\le r<400~\mathrm{nm},\\
H_B, & 400~\mathrm{nm}\le r<480~\mathrm{nm},\\
H_C, & r\ge480~\mathrm{nm}.
\end{cases}
\label{eq:segmented_field}
\end{equation}

When \(H_A=H_B\), their value is denoted by \(H_A=H_B\equiv H_{AB}\). In this case, regions A and B together form a circular uniform-field
region with a diameter of \(960~\mathrm{nm}\). Here, \(H_{AB}\) is a
shorthand notation for the common field in regions A and B rather than
an additional independent magnetic-field component. The quantities
\(H_A\), \(H_B\), and \(H_C\) are signed scalar components of the
perpendicular magnetic field. A positive field is directed along
\(+\hat{\mathbf z}\), whereas a negative field is directed along
\(-\hat{\mathbf z}\).

The proposed writing method is based on a field-gated nanocontact
protocol. Previous studies have shown that the size of skyrmions can
be tuned by a perpendicular magnetic field
\cite{tomasello2018origin,tejo2018distinct}. From the viewpoint of
Zeeman energy, a magnetic domain whose magnetization is parallel to
the applied field has a reduced energy and therefore tends to expand.
In contrast, a domain with magnetization antiparallel to the applied
field tends to shrink. By designing the spatial profile of the
magnetic field, either the outer bag or the inner bag can be
selectively expanded until the target region overlaps with the
nanocontact. In this sense, the segmented magnetic field serves as a
gate that determines which bag region is accessible to the
nanocontact. Once the target region reaches the nanocontact, a
spin-polarized current pulse is applied to create a local reversed
magnetic domain, which subsequently relaxes into a stable skyrmion
after the pulse is turned off. Since the outer and inner bags have
opposite magnetic backgrounds, current pulses with opposite effective
torque polarities are used to write skyrmions with opposite
topological charges in different bag regions.

The total micromagnetic energy of the system is written as
\begin{equation}
\begin{aligned}
E=\int dV \bigg[
& A(\nabla \mathbf{m})^2
+D\left(
m_z\nabla\cdot\mathbf{m}
-
(\mathbf{m}\cdot\nabla)m_z
\right)
-Km_z^2 \\
&-\mu_0 M_s H_z(\mathbf{r})m_z
-\frac{1}{2}\mu_0 M_s \mathbf{H}_{\rm d}\cdot\mathbf{m}
\bigg],
\end{aligned}
\label{eq:energy}
\end{equation}
where \(A\) is the exchange stiffness constant, \(D\) is the
interfacial DMI constant, \(K\) is the perpendicular magnetic
anisotropy constant, \(\mu_0\) is the vacuum permeability, \(M_s\) is
the saturation magnetization, \(H_z(\mathbf{r})\) is the spatially
dependent perpendicular magnetic field, and \(\mathbf{H}_{\mathrm d}\)
is the demagnetizing field. The spatially dependent field
\(H_z(\mathbf{r})\) accounts for the segmented magnetic field used to
control the expansion and contraction of different bag regions through
the Zeeman energy term.

The magnetization dynamics is governed by the
Landau--Lifshitz--Gilbert equation including a Slonczewski-type
spin-transfer torque
\cite{slonczewski1996current,sampaio2013nucleation,yuan2016skyrmion},
\begin{equation}
\frac{\partial \mathbf{m}}{\partial t}
=
-\gamma \mathbf{m}\times \mathbf{H}_{\mathrm{eff}}
+\alpha \mathbf{m}\times \frac{\partial \mathbf{m}}{\partial t}
+\boldsymbol{\tau}_{\mathrm{STT}},
\label{eq:llg}
\end{equation}
where \(\gamma\) is the gyromagnetic ratio, \(\alpha\) is the Gilbert
damping constant, and \(\mathbf{H}_{\mathrm{eff}}\) is the effective
magnetic field defined by
\begin{equation}
\mathbf{H}_{\mathrm{eff}}
=
-\frac{1}{\mu_0 M_s}
\frac{\delta E}{\delta \mathbf{m}}.
\label{eq:heff}
\end{equation}
The Slonczewski-type spin-transfer torque is given by
\begin{equation}
\boldsymbol{\tau}_{\mathrm{STT}}
=
-\gamma a_J \mathbf{m}\times
\left(
\mathbf{m}\times\mathbf{p}
\right),
\label{eq:stt}
\end{equation}
where \(\mathbf{p}\) is the spin-polarization direction of the injected
current. The coefficient \(a_J\) is written as
\begin{equation}
a_J=\frac{\hbar P J}{2eM_s t},
\label{eq:aj}
\end{equation}
where \(\hbar\) is the reduced Planck constant, \(P\) is the spin
polarization, \(J\) is the current density, \(e\) is the elementary
charge, and \(t\) is the thickness of the ferromagnetic layer. In the
simulations, the spin-polarization direction is fixed by the magnetization of the reference layer, \(\mathbf{p}=+\hat{\mathbf z}\), whereas reversing
the current direction changes the sign of the spin-transfer torque.
This allows magnetic textures with opposite core polarities to be
nucleated in the two oppositely magnetized local backgrounds.

The material parameters are based on those commonly used in
micromagnetic simulations of ultrathin Co-based films
\cite{sampaio2013nucleation}. We use
\(M_s=5.8\times10^5~\mathrm{A/m}\),
\(A=1.5\times10^{-11}~\mathrm{J/m}\),
\(K=8.0\times10^5~\mathrm{J/m^3}\),
\(D=3.5\times10^{-3}~\mathrm{J/m^2}\),
\(\alpha=0.3\), and \(P=0.4\). All micromagnetic simulations are performed using MuMax3 \cite{vansteenkiste2014design}. The
computational cell size is
\(1~\mathrm{nm}\times1~\mathrm{nm}\times1~\mathrm{nm}\), which is smaller than the magnetostatic exchange length \(l_{\rm ex}=
\sqrt{\frac{2A}{\mu_0M_s^2}}
\approx8.4~\mathrm{nm}\)\cite{abo2013definition}.

\begin{figure}
\centering
\includegraphics[width=8.5cm]{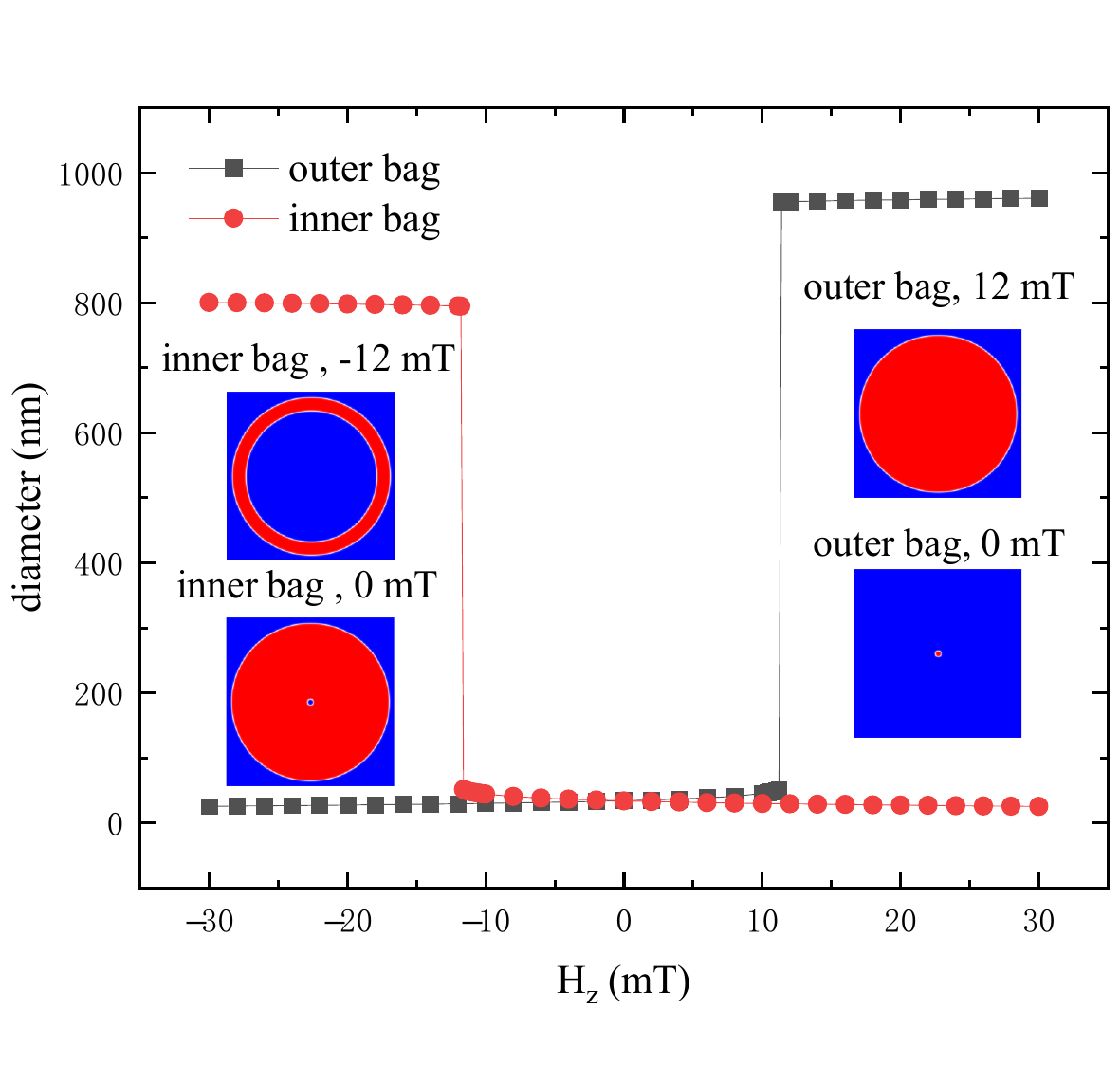}
\caption{
Field-dependent size modulation of the nested skyrmion bag. The field
in region C is fixed at \(H_C=-20~\mathrm{mT}\). The black curve shows
the outer bag diameter \(D_{\mathrm{out}}\) as a function of the common
field \(H_{AB}\), where \(H_A=H_B=H_{AB}\). The red curve shows the
inner bag diameter \(D_{\mathrm{in}}\) as a function of \(H_A\),
while the annular and outer fields are fixed at
\(H_B=+20~\mathrm{mT}\) and \(H_C=-20~\mathrm{mT}\), respectively.
}
\label{fig2}
\end{figure}

\section{Results and Discussion}
\subsection{Field-controlled size modulation and initialization}
We begin by examining the equilibrium configurations of the \(S(0)\)
skyrmion and the \(S(0,S(0))\) nested skyrmion bag under spatially
segmented perpendicular magnetic fields. For the equilibrium
calculations, the two textures are initialized using idealized
concentric reversed domains separated by N\'eel-type domain walls and
are subsequently relaxed at zero field using the conjugate-gradient
energy-minimization algorithm implemented in MuMax3. A perpendicular
magnetic field is then varied quasistatically, and the total
micromagnetic energy is minimized at each field value. Throughout the
calculation, the field in region C is fixed at
\(H_C=-20~\mathrm{mT}\). This oppositely oriented outer field confines
the outermost boundary within the field-gated region
(\(r<480~\mathrm{nm}\)) and suppresses uncontrolled radial expansion.

The size of each relaxed texture is extracted from the relevant closed
\(m_z=0\) contour. As shown in Fig.~\ref{fig2}, the black curve represents the outer
diameter \(D_{\mathrm{out}}\) of the isolated \(S(0)\) texture as a
function of the common field \(H_A=H_B=H_{AB}\). This texture
subsequently serves as the outer boundary of the dynamically
constructed \(S(0,S(0))\) state. The diameter
increases monotonically with \(H_{AB}\). For
\(H_{AB}<12~\mathrm{mT}\), the expansion is gradual, whereas a more
rapid increase occurs as \(H_{AB}\) approaches
\(12~\mathrm{mT}\). Above this geometric overlap threshold, the
nanocontact lies inside the expanded outer region, allowing
localized current-induced writing.

The red curve shows the inner-bag diameter \(D_{\mathrm{in}}\) as
a function of \(H_A\), with \(H_B=+20~\mathrm{mT}\) and
\(H_C=-20~\mathrm{mT}\). At \(H_A=H_B=+20~\mathrm{mT}\), the inner
bag remains contracted, while the outer bag is maintained in its
expanded state. Decreasing \(H_A\) toward the opposite polarity
selectively expands the inner bag. For
\(H_A\lesssim-12~\mathrm{mT}\), the inner region extends over the
nanocontact, whereas the expanded outer bag remains stabilized by the
fixed fields in regions B and C. Based on this behavior, we select the operational fields of \(H = \pm 20~\mathrm{mT}\) for the subsequent writing protocol, as these values ensure sufficient and stable expansion of the targeted bag region to reliably cover the nanocontact.

In addition to the idealized initialization used in the equilibrium
calculations, the \(S(0,S(0))\) state can be dynamically constructed
using two current-induced nucleation steps. A spin-polarized current pulse with \(I=+30~\mathrm{mA}\) and
\(\tau=50~\mathrm{ps}\) is first applied through the nanocontact,
nucleating an up-core \(S(0)\) skyrmion. The resulting skyrmion is then
enlarged by applying the appropriate \(H_{AB}\), creating a
sufficiently large reversed domain to accommodate a skyrmion with the
opposite core polarity. Subsequently, a current pulse of the opposite
polarity, \(I=-30~\mathrm{mA}\) and \(\tau=50~\mathrm{ps}\), is
applied through the nanocontact to nucleate a down-core skyrmion within
this expanded reversed domain. After relaxation, the resulting nested
configuration forms the \(S(0,S(0))\) skyrmion bag state.

\begin{figure*}[htbp]
\centering
\includegraphics[width=15cm]{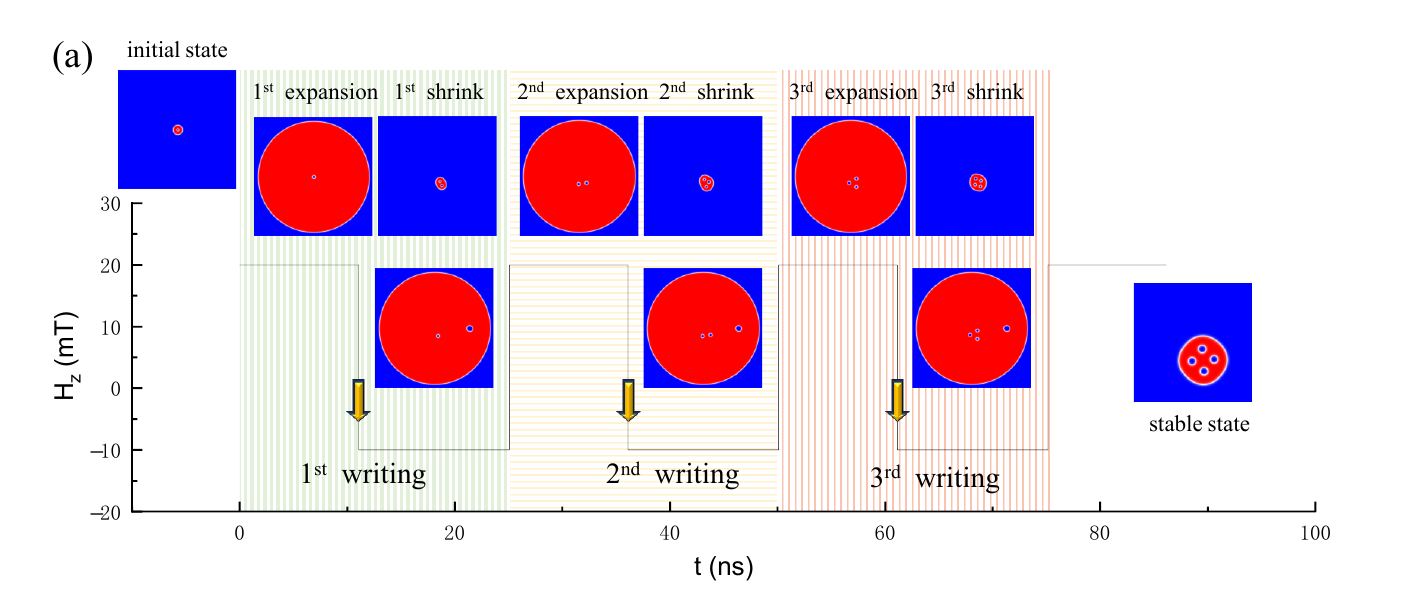}
\includegraphics[width=15cm]{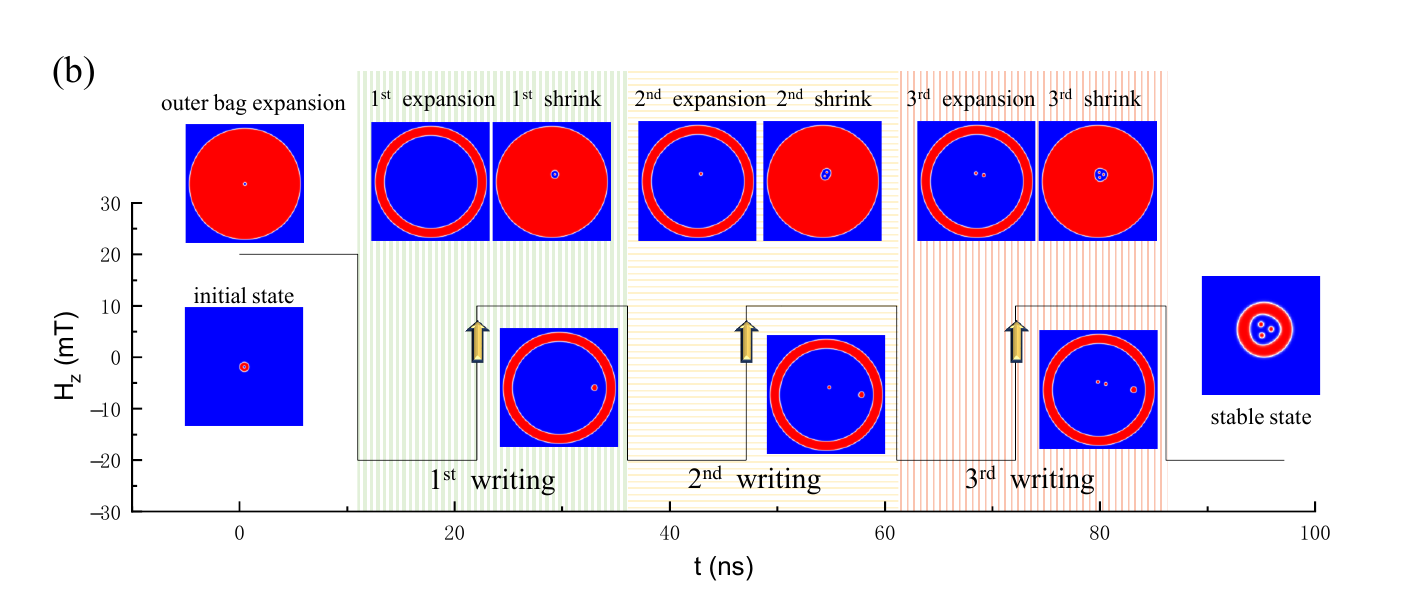}
\includegraphics[width=15cm]{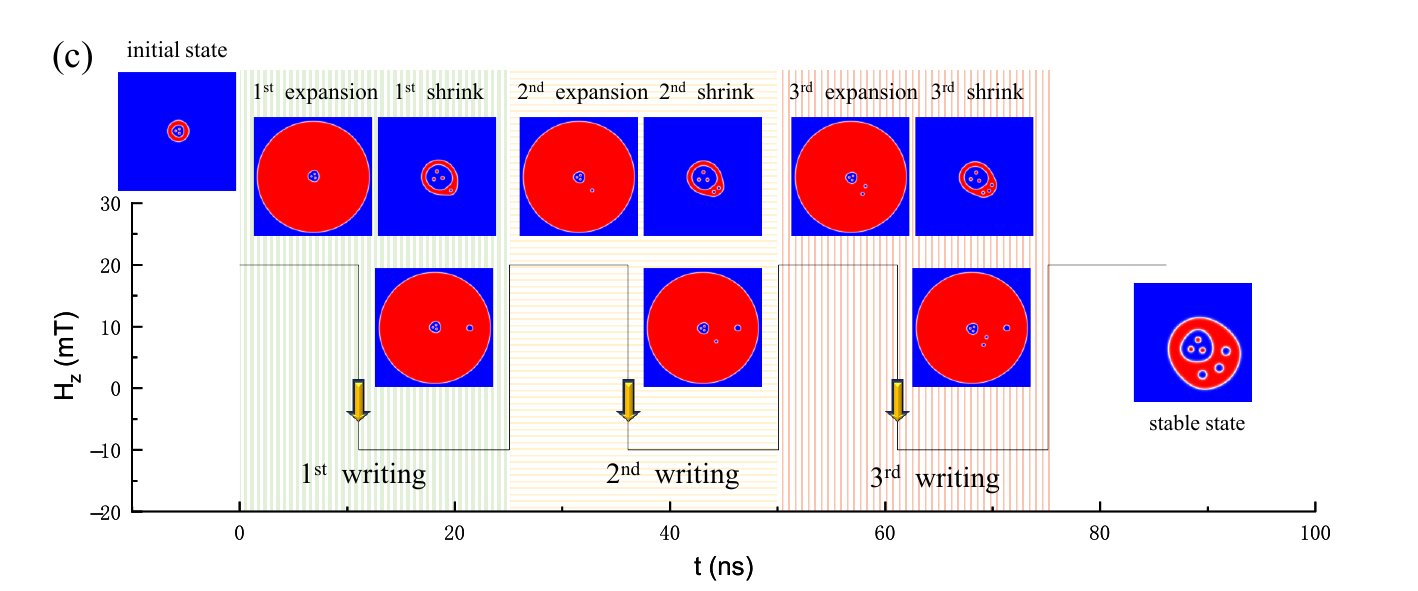}
\caption{
Sequential writing of nested skyrmion bags using the field-gated
nanocontact protocol. Colored background blocks indicate repeated
writing cycles consisting of field-driven expansion, current-induced
nucleation, and field-driven contraction. The upward- and
downward-pointing arrows denote current pulses of
\(I=+30~\mathrm{mA}\) and \(I=-30~\mathrm{mA}\), respectively, each
with a duration of \(50~\mathrm{ps}\). Positive current is defined as
flowing along \(+\hat{\mathbf z}\).
(a) Writing of the \(S(m,S(0))\) state. The curve represents the
variation of the magnetic field \(H_{AB}\). The outer bag is expanded
by setting \(H_A=H_B=H_{AB}=+20~\mathrm{mT}\) and contracted by setting
\(H_A=H_B=H_{AB}=-20~\mathrm{mT}\), while \(H_C\) is fixed at
\(-20~\mathrm{mT}\).
(b) Writing of the \(S(0,S(n))\) state. The curve represents the
variation of the magnetic field \(H_A\). The outer bag is first
expanded by setting \(H_A=H_B=H_{AB}=+20~\mathrm{mT}\). The inner
bag is then expanded by keeping \(H_B=+20~\mathrm{mT}\) fixed and
switching \(H_A\) to \(-20~\mathrm{mT}\). After writing, \(H_A\) is
restored to \(+20~\mathrm{mT}\) to contract the inner bag. During
all field-gated writing stages, the field in region C is maintained at
\(H_C=-20~\mathrm{mT}\).
(c) Sequential construction of an \(S(m,S(n))\) state by combining the
inner- and outer-region writing cycles. The curve represents the
variation of the magnetic field \(H_{AB}\).
}
\label{fig3}
\end{figure*}

\subsection{Region-selective writing of nested skyrmion bags}
The opposite field responses of the outer and inner regions
enable region-selective writing. In each writing cycle, the
segmented magnetic fields are first adjusted until the nanocontact
lies inside the target region. A \(50~\mathrm{ps}\)
spin-polarized current pulse is then applied to nucleate a skyrmion
against the local background magnetization. Finally, the target
region is contracted, thereby transporting the newly nucleated
skyrmion away from the nanocontact and restoring the writing region
for the next cycle.

Because the outer and inner regions have opposite background
magnetizations, opposite current polarities are required. Under the
current convention adopted here, the outer and inner writing pulses
are \(I_{\mathrm{out}}=-30~\mathrm{mA}\) and
\(I_{\mathrm{in}}=30~\mathrm{mA}\), respectively. The two types of
nucleated skyrmions consequently carry opposite topological charges.

First, we demonstrate the generation of the \(S(m,S(0))\) state from
the initial \(S(0,S(0))\) configuration, as shown in Fig.~\ref{fig3}(a). To write a skyrmion into the outer region, the
fields in regions A and B are first set to the same value,
\(H_A=H_B=H_{AB}=+20~\mathrm{mT},\)
while the field in region C is kept fixed at
\(H_C=-20~\mathrm{mT}\).
These fields are applied for \(11~\mathrm{ns}\). The positive
\(H_{AB}\) drives the expansion of the outer bag toward the
nanocontact, whereas the negative \(H_C\) suppresses further outward
expansion once the outer boundary approaches region C. As a result,
the outer bag reaches a stable expanded state and overlaps with the
nanocontact. While maintaining \(H_{AB}=+20~\mathrm{mT}\) and
\(H_C=-20~\mathrm{mT}\), a spin-polarized current pulse is applied
through the nanocontact to nucleate one skyrmion with topological
charge \(Q=-1\) in the outer region. Because the newly generated
skyrmion is initially located near the nanocontact, the outer bag is
subsequently contracted to move it away from the writing region. For
this purpose, the common field in regions A and B is reversed to
\(H_A=H_B=H_{AB}=-20~\mathrm{mT},\)
while \(H_C\) remains fixed at \(-20~\mathrm{mT}\). These fields are
applied for \(14~\mathrm{ns}\), causing the outer bag to contract and
carry the newly generated skyrmion away from the nanocontact. The
skyrmion remains confined within the outer region after the bag
returns to a stable size. Repeating this
expansion--injection--contraction cycle \(m\) times yields the
\(S(m,S(0))\) state.

Second, we demonstrate the generation of \(S(0,S(n))\), in which the
inner bag is selected as the writing target, as shown in
Fig.~\ref{fig3}(b). Because the inner bag does not
initially overlap with the nanocontact, the outer bag is first
expanded by setting
\(H_A=H_B=H_{AB}=+20~\mathrm{mT},\)
while the outer field is maintained at
\(H_C=-20~\mathrm{mT}.\)
These fields are applied for \(11~\mathrm{ns}\), allowing the outer
bag to expand to the required state. After the outer bag reaches the
expanded state, the annular field is kept fixed at
\(H_B=+20~\mathrm{mT},\)
whereas the central field \(H_A\) is switched from \(+20~\mathrm{mT}\) to
\(-20~\mathrm{mT}.\)
The field \(H_C\) remains fixed at \(-20~\mathrm{mT}\). This field
configuration is maintained for \(11~\mathrm{ns}\). The positive
\(H_B\) stabilizes the expanded outer bag, whereas the negative
\(H_A\) selectively drives the expansion of the inner bag toward
the nanocontact. Once the inner bag overlaps with the nanocontact,
a current pulse with the opposite polarity to that used for
outer-region writing is applied. This pulse nucleates one
skyrmion with topological charge \(Q=+1\) in the inner bag. After
writing, \(H_B\) is kept fixed at \(+20~\mathrm{mT}\), while \(H_A\)
is switched back from \(-20~\mathrm{mT}\) to
\(+20~\mathrm{mT}\). The fields in regions A and B then become
identical again and can be written as
\(H_A=H_B=H_{AB}=+20~\mathrm{mT}.\)
Meanwhile, \(H_C\) remains fixed at \(-20~\mathrm{mT}\). The reversal
of \(H_A\) causes the inner bag to contract and move the newly
generated skyrmion away from the nanocontact, while \(H_B\) maintains
the expanded outer bag. Repeating this inner-bag
expansion--injection--contraction cycle \(n\) times yields the
\(S(0,S(n))\) state.

Finally, a general \(S(m,S(n))\) state is constructed by combining the inner- and outer-region writing procedures, as shown in Fig.~\ref{fig3}(c). First, \(n\)
skyrmions with topological charge \(Q=+1\) are sequentially written
into the inner bag using the inner-bag writing cycle described
above, thereby forming an \(S(0,S(n))\) state. Starting from this
preformed state, skyrmions are subsequently written into the outer
region. The outer bag is first expanded by setting
\(H_A=H_B=H_{AB}=+20~\mathrm{mT}\), while \(H_C\) is maintained at \(-20~\mathrm{mT}\). After the outer bag overlaps with the nanocontact, a current pulse is applied to nucleate one skyrmion with topological charge \(Q=-1\) in the outer
region. The common field in regions A and B is then reversed to
\(H_A=H_B=H_{AB}=-20~\mathrm{mT},\)
while \(H_C\) remains fixed at \(-20~\mathrm{mT}\). This change
contracts the outer bag and moves the newly generated skyrmion away
from the nanocontact. Repeating the outer-region writing cycle
\(m\) times yields the desired \(S(m,S(n))\) state.

\begin{figure}
\centering
\includegraphics[width=8.5cm]{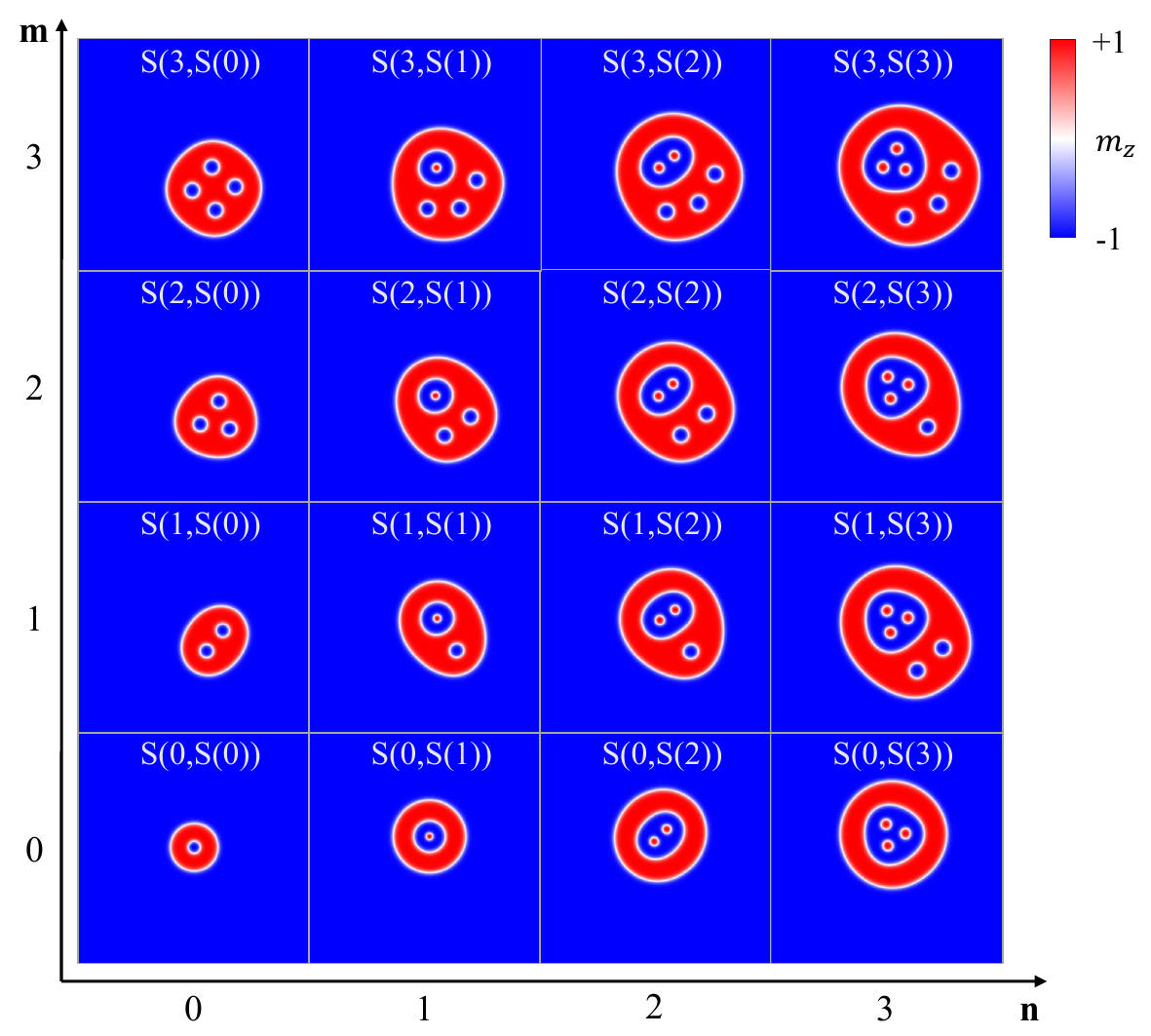}
\caption{
Representative magnetization configurations of double-layer nested skyrmion bags \(S(m,S(n))\) with different numbers of skyrmions in the outer and inner bag regions. Each panel shows a central
\(512~\mathrm{nm}\times512~\mathrm{nm}\) field of view cropped from
the full \(1024~\mathrm{nm}\times1024~\mathrm{nm}\) simulation domain. The vertical and horizontal axes denote the numbers of skyrmions in the outer bag, \(m\), and in the inner bag, \(n\), respectively.
}
\label{fig4}
\end{figure}
\begin{figure*}[!t]
\centering
\includegraphics[width=5.1cm]{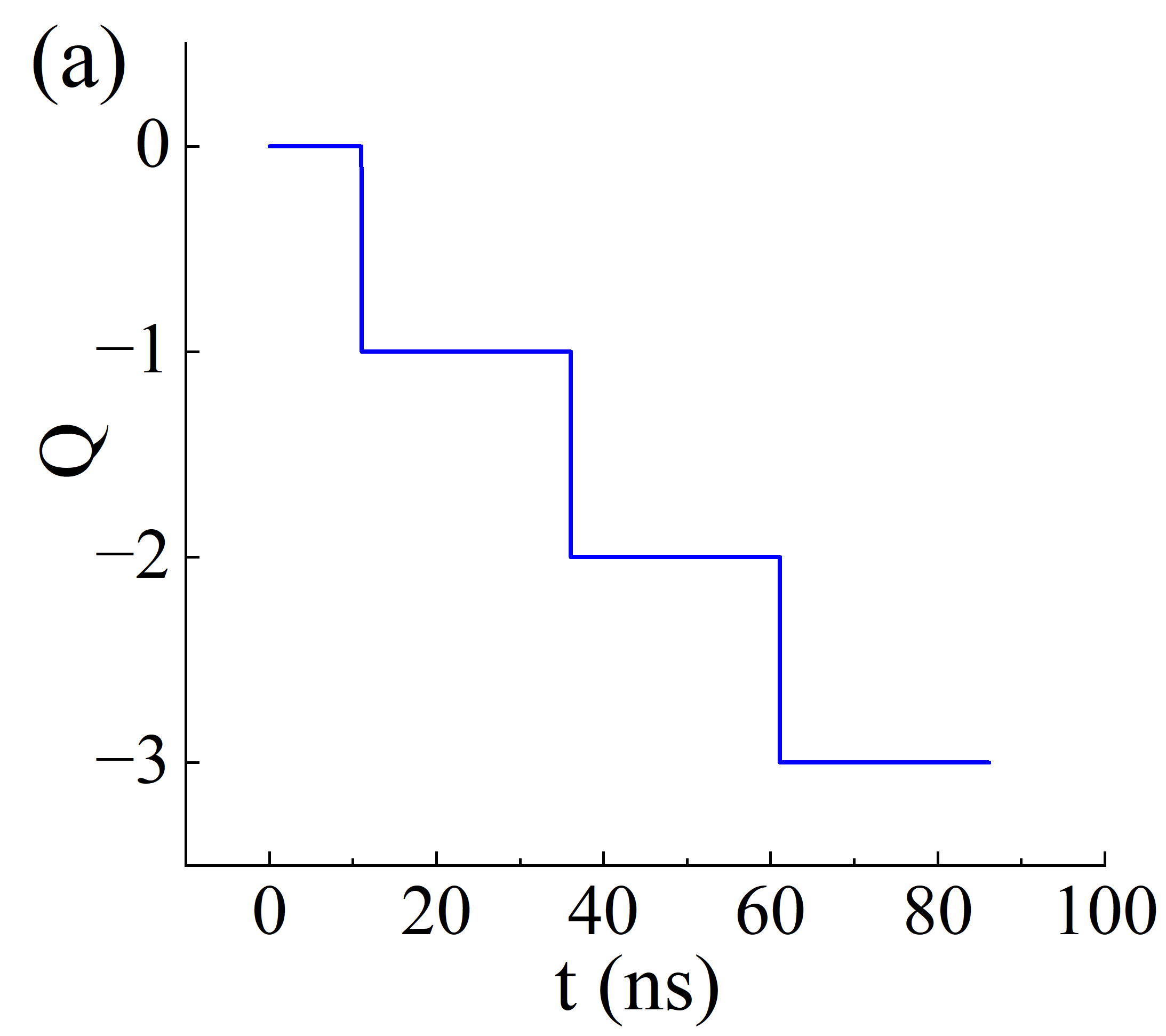}
\includegraphics[width=5.1cm]{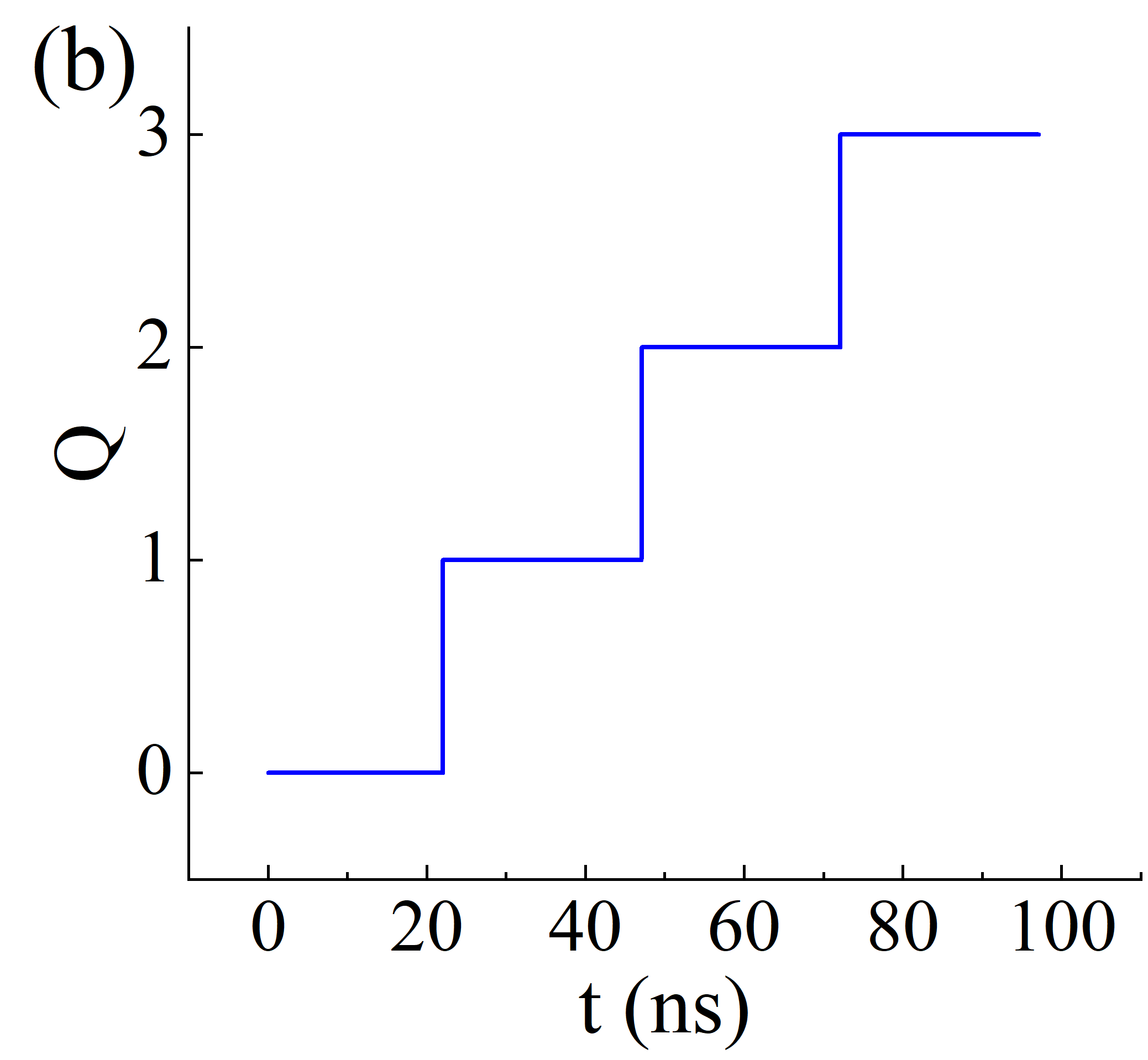}
\includegraphics[width=5.1cm]{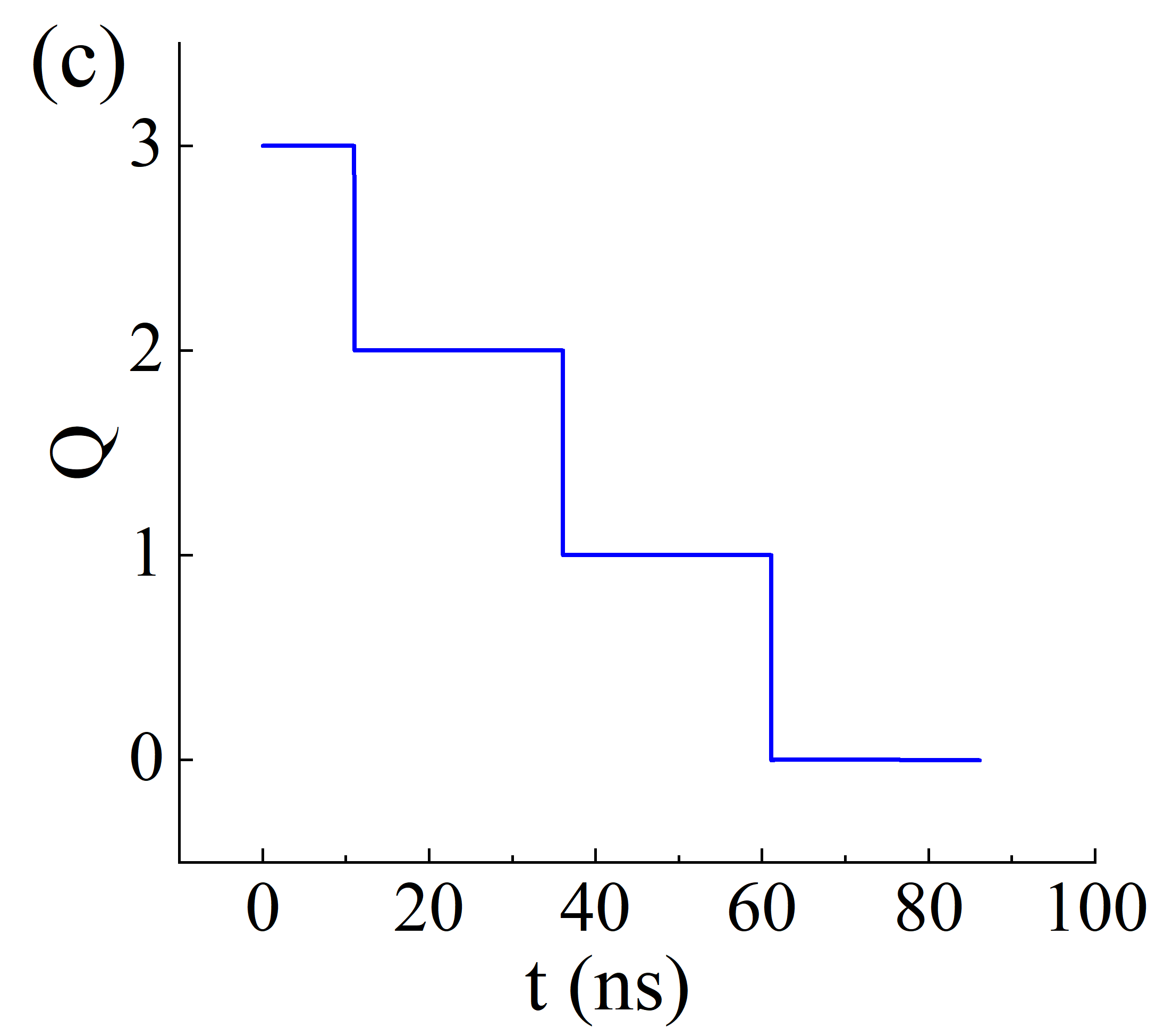}
\caption{
Time evolution of the total topological charge \(Q_{\mathrm{total}}\) during the writing processes of nested skyrmion bags.
(a) Sequential generation of \(S(3,S(0))\), where \(Q_{\mathrm{total}}\) decreases from \(0\) to \(-3\).
(b) Sequential generation of \(S(0,S(3))\), where \(Q_{\mathrm{total}}\) increases from \(0\) to \(+3\).
(c) Sequential generation of \(S(3,S(3))\) from a preformed \(S(0,S(3))\) state, where \(Q_{\mathrm{total}}\) decreases from \(+3\) to \(0\).
The stepwise variation confirms that each writing cycle changes the total topological charge by one unit.
}
\label{fig5}
\end{figure*}
\subsection{Programmable states and topological-charge evolution}
Figure~\ref{fig4} shows representative magnetization configurations
of double-layer nested skyrmion bags \(S(m,S(n))\) with different values of
\(m\) and \(n\). Here, \(m\) denotes the number of skyrmions confined in the
outer bag, while \(n\) denotes the number of skyrmions confined in the inner
bag. The vertical and horizontal directions in the configuration map
therefore correspond to increasing \(m\) and \(n\), respectively.

Because the outer and inner bag regions have opposite background
magnetizations, skyrmions written into these two regions contribute opposite
unit topological charges. Under the present convention, each skyrmion in the
outer bag carries \(Q=-1\), whereas each skyrmion in the inner bag carries
\(Q=+1\). According to Eq.~\eqref{eq:qtotal}, the total topological charge is determined by the difference between the two occupation numbers.
This relation is directly reflected in the configuration map. Increasing
\(n\) along the horizontal direction increases \(Q_{\mathrm{total}}\), whereas
increasing \(m\) along the vertical direction decreases
\(Q_{\mathrm{total}}\). For example, \(S(3,S(0))\) has
\(Q_{\mathrm{total}}=-3\), \(S(0,S(3))\) has \(Q_{\mathrm{total}}=+3\), and
\(S(3,S(3))\) has \(Q_{\mathrm{total}}=0\).

It is worth noting that the configurations along the \(m=n\) diagonal, such as
\(S(1,S(1))\), \(S(2,S(2))\), and \(S(3,S(3))\), have a compensated total
topological charge \(Q_{\mathrm{total}}=0\), although they still contain
multiple skyrmionic objects inside the bag. This feature indicates that
nested skyrmion bags provide a flexible platform for programming positive,
negative, and zero total topological charges by independently controlling the
numbers of skyrmions written into the outer and inner bag regions.

To further verify the controllability of the proposed writing protocol, we
calculate the time evolution of the total topological charge during the
generation of representative nested skyrmion-bag states, as shown in
Fig.~\ref{fig5}. For the writing of \(S(3,S(0))\), the total
topological charge decreases stepwise from \(0\) to \(-3\)
[Fig.~\ref{fig5}(a)]. Each decrease by one unit corresponds to the
nucleation of one skyrmion in the outer bag, which carries a unit topological
charge \(Q=-1\). The plateau after each abrupt change indicates that the
newly written skyrmion remains stable during the subsequent relaxation and
field-driven contraction processes.

In contrast, when skyrmions are written into the inner bag, the total
topological charge increases stepwise from \(0\) to \(+3\), as shown in
Fig.~\ref{fig5}(b). This behavior is consistent with the opposite
magnetic background of the inner bag, where each injected skyrmion carries
\(Q=+1\). Therefore, the sign of the topological-charge variation can be controlled by selecting the target bag region.

Figure~\ref{fig5}(c) shows the sequential construction of the
compensated \(S(3,S(3))\) state. Starting from the preformed \(S(0,S(3))\) configuration with \(Q_{\mathrm{total}}=+3\), three skyrmions
are subsequently written into the outer bag. As a result, \(Q_{\mathrm{total}}\) decreases stepwise from \(+3\) to \(0\). The final state has zero total topological charge because the positive contribution from the three inner-bag skyrmions is exactly compensated by the negative contribution from the three outer-bag skyrmions. The stepwise variation confirms that each successful nucleation event
produces a net change of approximately one unit in the total
topological charge, while the subsequent field-driven deformation and
relaxation preserve the topology. These results demonstrate that the proposed field-gated nanocontact protocol enables deterministic control of the total topological charge of nested skyrmion bags.

\section{Conclusion and Outlook}

In conclusion, we have proposed a field-gated nanocontact protocol for
the controlled generation and region-selective writing of double-layer nested skyrmion bags \(S(m,S(n))\). By using spatially segmented
perpendicular magnetic fields, the outer region or inner bag
can be selectively expanded to overlap with the nanocontact. A
spin-polarized current pulse then nucleates one skyrmion in the target
region, and the subsequent contraction restores the writing site
for repeated operation. This protocol enables the construction of
\(S(m,S(0))\), \(S(0,S(n))\), and general \(S(m,S(n))\) states.

Because skyrmions in the two regions contribute opposite
topological charges, the total topological charge follows
\(Q_{\mathrm{total}} = n-m\). Its stepwise evolution confirms
one-at-a-time and region-selective skyrmion writing, allowing
positive, negative, and compensated topological-charge states to be
programmed. Moreover, states with the same total topological charge
can have different occupation pairs \((m,n)\), providing an additional
configurational degree of freedom for multilevel information encoding.

Future studies should examine the effects of temperature, material
disorder, smooth field boundaries, and current-induced heating on the
writing reliability and stability of these states. Experimental
realization of local magnetic gating and electrical readout of the two
skyrmion populations will also be important. These results provide a
possible route toward programmable nested topological states and
multistate spintronic devices.

\section*{Acknowledgments}
This work was supported by the National Natural Science Foundation of China (Grant No. 12104296) and the startup funding from Great Bay
University (No. YJKY220018).

\bibliography{apssamp}

\end{document}